\documentclass[pra,showpacs,twocolumn]{revtex4}%
\usepackage{amsfonts}
\usepackage{amsmath}
\usepackage{amssymb,amscd}
\usepackage{psfrag}
\usepackage{graphicx}
\usepackage{braket}
\usepackage[dvips]{epsfig}
\usepackage{epsfig}
\usepackage{subfigure}
\usepackage{color}
\usepackage{amssymb}%
\providecommand{\U}[1]{\protect\rule{.1in}{.1in}}
\begin{document}
\title{Intramode correlations enhanced phase sensitivities in an SU(1,1) interferometer}
\author{Qian-Kun Gong$^{1}$, Dong Li$^{1}$, Chun-Hua Yuan$^{1,4}$}
\email{chyuan@phy.ecnu.edu.cn}
\author{Z. Y. Ou$^{1,3}$}
\author{Weiping Zhang$^{2,4}$}

\address{$^1$Quantum Institute for Light and Atoms, Department of Physics, East China
Normal University, Shanghai 200062, P. R. China}

\address{$^2$ Department of Physics and Astronomy, Shanghai Jiao Tong University, Shanghai 200240, P. R. China}

\address{$^{3}$Department of Physics, Indiana University-Purdue University
Indianapolis, 402 North Blackford Street, Indianapolis, Indiana 46202, USA}

\address{$^{4}$Collaborative Innovation Center of Extreme Optics, Shanxi University, Taiyuan, Shanxi 030006, P. R. China}

\begin{abstract}
We theoretically derive the lower and upper bounds of quantum Fisher
information (QFI) of an SU(1,1) interferometer whatever the input state
chosen. According to the QFI, the crucial resource for quantum enhancement is
shown to be large intramode correlations indicated by the Mandel
$Q$-parameter. For a photon-subtracted squeezed vacuum state with high
super-Poissonian statistics in one input port and a coherent state in the
other input port, the quantum Cram\'{e}r-Rao bound of the SU(1,1)
interferometer can beat $1/\langle\hat{N}\rangle$ scaling in presence of large
fluctuations in the number of photons, with a given fixed input mean number of
photons. The definition of the Heisenberg limit (HL) should take into account
the amount of fluctuations. The HL considering the number fluctuation effect
may be the ultimate phase limit.

\end{abstract}
\date{\today}

\pacs{42.50.St, 07.60.Ly, 42.50.Lc, 42.65.Yj}
\maketitle

\section{Introduction}

Quantum mechanics provides a fundamental limit on the achievable measurement
precision, optimized over all possible estimators, measurements and probe
states. The paradigmatic example is the optical phase estimation. When the
number of particles in the input state is fixed and equal to $N$, the phase
sensitivity is limited by two bounds. One is the standard quantum limit (SQL),
$1/\sqrt{N}$, which is due to the classical nature of the coherent state and
can be beaten by quantum-mechanical effects. Another is the fundamental
Heisenberg limit (HL) which is given by $1/N$ and cannot be beaten.
Analogously, for a fixed mean photon number $\langle\hat{N}\rangle$,
measurement accuracy also has a fundamental HL. However, the violation of
$1/\langle\hat{N}\rangle$ scaling of the optimal accuracy is possible in
presence of large fluctuations in the number of probes~\cite{Sharpio,Pasquale}%
. Then definition of the HL should take into account the amount of photon
number fluctuations~\cite{hofmann,hyllus,pezze2015}.

Using quantum measurement techniques to beat the SQL, has been received a lot
of attention in recent years
\cite{Helstrom67,Holevo82,Caves81,Braunstein94,Braunstein96,Lee02,Giovannetti06,Zwierz10,Giovannetti04,Giovannetti11}%
. The Mach-Zehnder interferometer (MZI) and its variants, which can be
understood as a two-mode (two-path) interferometer, have been used as a
generic model to realize precise measurement of phase \cite{Abbott}. In order
to avoid the vacuum fluctuations, Caves suggested to use the coherent and
squeezed-vacuum light as input of a Mach-Zehnder interferometer to reach a
sub-shot-noise sensitivity in 1981 \cite{Caves81}. Later, many quantum
parameter estimation protocols have been proposed
\cite{Giovannetti04,Giovannetti11,Toth}. The enhancements obtained from
employing quantum state can be divided into two parts according to the
correlations \cite{Sahota}: (1) intramode correlations which are provided by a
large uncertainty in the photon number in each arm, such as the squeezed
vacuum exhibits high intramode correlations due to nonclassical photon
statistics; (2) intermode correlations, i.e., correlations between two paths
which can be realized by mode entanglement. The NOON state, i.e., states of
the form $(|N\rangle_{a}|0\rangle_{b}+e^{i\phi_{N}}|0\rangle_{a}|N\rangle
_{b})/\sqrt{2}$ have been suggested to reach the HL scaling, in the
phase-shift measurements $\Delta\phi_{\text{HL}}=1/N$ \cite{NOON,Dowling}. The
enhancement of phase sensitivity with NOON state benefits from both intermode
correlation and intramode correlation \cite{Knott}. However, the intermode
correlations can contribute at most a factor of $1/\sqrt{2}$ improvement in
the phase precision, which was pointed out by Sahota and Quesada
\cite{Sahota}. But the intramode correlations have no upper bound
\cite{Berrada}, which naturally leads us to search and study quantum states
with high intramode correlations.

\begin{figure}[ptbh]
\centerline{\includegraphics[scale=0.62,angle=0]{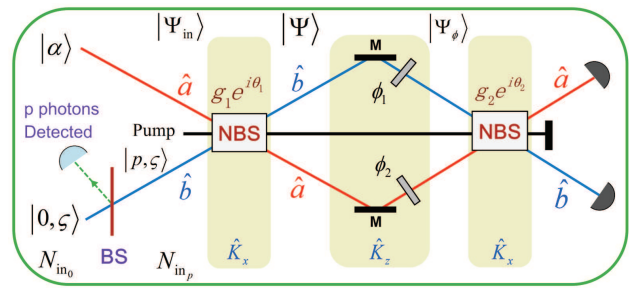}}\caption{(Color
online) The schematic diagram of the SU(1,1) interferometer. Two Nonlinear
beam splitters (NBSs) take the place of two beam splitters in the traditional
Mach-Zehnder interferometer, and $\hat{K}_{i}$ $(i=x,y,z)$ instead of $\hat
{J}_{i}$ $(i=x,y,z)$ describe this interferometer in Schwinger representation.
The input state $|\Psi_{\text{in}}\rangle$ injecting into the first NBS leads
to output state $|\Psi\rangle$, then it is modified as $|\Psi_{\phi}\rangle$
by phase shift. $g_{1}$ ($g_{2}$) and $\theta_{1}$ ($\theta_{2}$) describe the
strength and phase shift in the NBS process $1$ ($2$), respectively. $\hat{a}$
and $\hat{b}$ denote two light modes in the interferometer. A coherent state
$|\alpha\rangle$ is in one input port and a squeezed-vacuum state
$|0,\varsigma\rangle$ in the other input port of the SU(1,1) interferometer
with total mean photon number of input $N_{\text{in}_{0}}$. A beam splitter
with low reflectance and a single-photon resolution photo-detector are used to
subtract $p$-photons from input squeezed vacuum state. The reflected photons
that are detected herald a $p$-photon subtracted squeezed-vacuum state
generation $|p,\varsigma\rangle_{b}$ $(\sim\hat{b}^{p}|0,\varsigma\rangle
_{b})$. For a coherent light combined with a $p$-photon subtracted squeezed
vacuum light as input, the total mean photon number of input is $N_{\text{in}%
_{p}}$. $\phi_{1}$, $\phi_{2}$: phase shift; M: mirrors; BS: beam splitter.}%
\label{fig1}%
\end{figure}

In additional to the nonclassical input states with high intramode
correlations, the nonlinear elements were also introduced in the linear
interferometers to improve the measurement precision. Such a class of
interferometers introduced by Yurke \emph{et al.} \cite{Yurke86} is described
by the group SU(1,1)-as opposed to SU(2), where the 50-50 beam splitters (BSs)
in a traditional MZI was replaced by the nonlinear beam splitters (NBSs), such
as optical parametric amplifiers (OPAs) or four-wave mixings (FWMs) (see
Fig.~\ref{fig1}). Recently, an improved theoretical scheme was presented by
Plick \emph{et al.} \cite{Plick} who proposed to inject a strong coherent beam
to \textquotedblleft boost\textquotedblright\ the photon number. With a
coherent state in one input port and a squeezed-vacuum state in the other
input port using the method of homodyne detection, the phase sensitivity can
approach $1/\langle\hat{N}\rangle$ scaling \cite{Li14}. Experimental
realization of this SU(1,1) optical interferometer was reported by our
group~\cite{Jing11}. The noise performance of this interferometer was
analyzed~\cite{Ou,Marino} and under the same phase-sensing intensity condition
the improvement of $4.1$ dB in signal-to-noise ratio was
observed~\cite{Hudelist}. By contrast, an SU(1,1) atomic interferometer has
also been experimentally realized with Bose-Einstein
Condensates~\cite{Gross,Linnemann,Peise,Gabbrielli}. An atom-photon interface
can form an atom-light hybrid interferometer
\cite{Yama,Haine,Szigeti,Haine16,ChenPRL15}, where the atomic Raman
amplification processes take the place of the beam splitting elements in a
traditional MZI, and the losses performance of it was analyzed recently
\cite{Yuan16}. Different from all-optical or all-atomic interferometers, the
main feature is that both the optical field and atomic phases can be probed
with optical interferometric techniques. In addition, the circuit quantum
electrodynamics system was also introduced to form the SU(1,1) interferometer
\cite{Barzanjeh}, which provides a different method for basic measurement
using the hybrid interferometers.

Recently, Lang and Caves \cite{Lang} have proved that if one input port
injected by a coherent state in an MZI, in the other input port the best
injection\ state is a squeezed vacuum state with a given fixed mean number of
photons. However, since the subtraction of photons from squeezed vacuum state
has the effect of increasing the average photon number of the new field state,
as well as the intramode correlations \cite{Biswas}, Birrittella and Gerry
\cite{Gerry} suggested to use coherent and photon-subtracted squeezed vacuum
states as input for quantum optical interferometry, and they found that it was
still possible to attain higher sensitivity via photon subtraction from the
squeezed state. The fact that photon subtraction via the annihilation operator
leads to the average photon number counter-intuitively increasing was studied
over twenty-years ago \cite{Ueda, Mizrahi}. Ueda \textit{et al.} \cite{Ueda}
studied the average and variance of the field acted by the annihilation
operator. The difference of average photon number between the remaining field
$\bar{N}_{\text{sub}}$ and the initial field $\bar{N}$ is equivalent to the
Mandel's $Q$-parameter \cite{Mandel}, i.e., $\bar{N}_{\text{sub}}-\bar{N}=Q$.
For $Q>0$ ($-1<Q<0$), the photon statistics is super-Poissonian
(sub-Poissonian), then the average photon number of the remaining field is
shown to increase (decrease). For the Poissonian state (e.g., coherent light),
the average photon number does not change ($\bar{N}_{\text{sub}}=\bar{N}$).
Especially, due to generation of Schr\"{o}dinger cat states the
photon-subtracted squeezed vacuum states have been intensely studied
theoretically \cite{Biswas,Dakna,Takeola} and experimentally
\cite{Ourjoumtsev,Wakui,Namekata,Gerrits}. Up to now three-photon subtraction
from a squeezed vacuum state was reported \cite{Gerrits}.

In this paper, we theoretically derive the lower and upper bounds of quantum
Fisher information (QFI) \cite{Braunstein94,PezzBook} of an SU(1,1)
interferometer, in which the dominant resource for quantum enhancement is the
large intramode correlations indicated by the Mandel $Q$-parameter.
Furthermore, we study the phase sensitivities of the SU(1,1) interferometer
for a coherent light combined with a photon-subtracted squeezed vacuum light
and compare them with the HL.

Our article is organized as follows. In Sec. II, we give the QFI of the the
SU(1,1) interferometer for general pure states input, and show that the
metrological advantage of nonclassical light is primary the intramode
correlations (high Mandel's $Q$-parameter). Then we derive the QFI and give
the quantum Cram\'{e}r-Rao bound (QCRB) \cite{Helstrom67,Holevo82} of SU(1,1)
interferometer with a coherent state $\otimes$ photon-subtracted squeezed
vacuum state input. The comparison between the QCRB and the HL is given in
Sec. III. In Sec. IV, we compare our proposal with that of Lang and Caves
\cite{Lang}. Finally, we conclude with a summary of our results.

\section{Quantum Fisher information}

In general, it is difficult to optimize over the detection methods to obtain
the optimal estimation protocols. One of the common ways to obtain the lower
bounds in quantum metrology, is to use the method of the QFI. The so-called
QFI is defined by maximizing the Fisher information over all possible
measurement strategies allowed by quantum mechanics. It characterizes the
maximum amount of information that can be extracted from quantum experiments
about an unknown parameter using the best (and ideal) measurement device. It
establishes the best precision that can be attained with a given quantum probe.

\subsection{General description}

An SU(1,1) interferometer is shown in Fig.~\ref{fig1}, where NBSs replace the
50-50 BSs in a traditional MZI. We firstly give a brief review of the SU(1,1)
interferometer introduced by Yurke \emph{et al.} \cite{Yurke86}. The
traditional MZI is called an SU(2) interferometer and can be understood as a
two-mode (two-path) interferometer. The transformation by this interferometer
is a rotation on angular momenta observables $\hat{J}_{x}=\hbar(\hat
{a}^{\dagger}\hat{b}+\hat{a}\hat{b}^{\dagger})/2$, $\hat{J}_{y}=-i\hbar
(\hat{a}^{\dagger}\hat{b}-\hat{a}\hat{b}^{\dagger})/2$, and $\hat{J}_{z}%
=\hbar(\hat{a}^{\dagger}\hat{a}-\hat{b}^{\dagger}\hat{b})/2$, where $\hat{a}$
$(\hat{a}^{\dag})$ and $\hat{b}$ $(\hat{b}^{\dag})$ are the annihilation
(creation) operators corresponding to the two modes $a$ and $b$, respectively.
For the SU(1,1) interferometer, the induced transformation on the input
observables is that of the group SU(1,1). The Hermitian operators $\hat{K}%
_{x}=\hbar(\hat{a}^{\dagger}\hat{b}^{\dagger}+\hat{a}\hat{b})/2$, $\hat{K}%
_{y}=-i\hbar(\hat{a}^{\dagger}\hat{b}^{\dagger}-\hat{a}\hat{b})/2$, and
$\hat{K}_{z}=\hbar(\hat{a}^{\dagger}\hat{a}+\hat{b}^{\dagger}\hat{b}+1)/2$ are
introduced to describe the SU(1,1) interferometer. The initial state
$|\Psi_{\text{in}}\rangle$ injecting into a NBS results in the output
$|\Psi\rangle=e^{-i\xi\hat{K}_{x}}|\Psi_{\text{in}}\rangle$ where $\xi$ is the
two-mode squeezed parameter of the NBS \cite{TMS}. After the first NBS, the
two beams sustain phase shifts, i.e., mode $a$ undergoes a phase shift of
$\phi_{1}$\ and mode $b$ undergoes a phase shift of $\phi_{2}$. In the
Schr\"{o}dinger picture the state vector is transformed as $|\Psi_{\phi
}\rangle=e^{-i\phi\hat{K}_{z}}\left\vert \Psi\right\rangle $, where
$\phi=-(\phi_{1}+\phi_{2})$ \cite{Kok}.

The Fisher information (FI) $F$ is determined by the measurement statistics
used to estimation $\phi$. If a positive operator-valued measure (POVM)
$\{\hat{\Pi}_{i}\}$ describes a measurement on the modified probe state
$|\Psi_{\phi}\rangle$, then the FI $F$ is given by \cite{Kok,PezzBook}
\begin{equation}
F=\sum_{i}\frac{1}{\langle\Psi_{\phi}|\hat{\Pi}_{i}|\Psi_{\phi}\rangle}\left(
\frac{\partial\langle\Psi_{\phi}|\hat{\Pi}_{i}|\Psi_{\phi}\rangle}%
{\partial\phi}\right)  ^{2}.
\end{equation}
Various observables may lead to different $F$. Fortunately, the QFI is the
intrinsic information in the quantum state and is not related to actual
measurement procedure. The QFI is at least as great as the FI for the optimal
observable. To obtain the sensitivity of phase-shift measurements with our
input states, we can use the QFI to determine the maximum level of sensitivity
by the QCRB. The QFI $\mathcal{F}$ for a pure state is given by
\cite{Jarzyna,PezzBook}%
\begin{equation}
\mathcal{F}=4(\langle\Psi_{\phi}^{\prime}|\Psi_{\phi}^{\prime}\rangle
-\left\vert \langle\Psi_{\phi}^{\prime}|\Psi_{\phi}\rangle\right\vert ^{2}),
\end{equation}
where $|\Psi_{\phi}\rangle$ is the state vector just before the second NBS\ of
the SU(1,1) interferometer and $|\Psi_{\phi}^{\prime}\rangle=\partial
|\Psi_{\phi}\rangle/\partial\phi=-i\hat{K}_{z}\left\vert \Psi_{\phi
}\right\rangle $. Then the QFI is given by%
\begin{equation}
\mathcal{F}=4\Delta^{2}\hat{K}_{z},
\end{equation}
where $\Delta^{2}\hat{K}_{z}=\langle\Psi|\hat{K}_{z}^{2}|\Psi\rangle
-\langle\Psi|\hat{K}_{z}|\Psi\rangle^{2}$. As is described by two-mode model
and shown in Fig.~\ref{fig1}, the QFI $\mathcal{F}$ can be written as%
\begin{equation}
\mathcal{F}=\Delta^{2}\hat{n}_{a}+\Delta^{2}\hat{n}_{b}+2Cov[\hat{n}_{a}%
,\hat{n}_{b}], \label{Fisher}%
\end{equation}
where $\hat{n}_{a}=\hat{a}^{\dag}\hat{a}$, $\hat{n}_{b}=\hat{b}^{\dag}\hat{b}%
$, $\Delta^{2}\hat{n}_{i}=\langle\Psi|\hat{n}_{i}^{2}|\Psi\rangle-\langle
\Psi|\hat{n}_{i}|\Psi\rangle^{2}$ $(i=a$, $b)$, and $Cov[\hat{n}_{a},\hat
{n}_{b}]=\langle\Psi|\hat{n}_{a}\hat{n}_{b}|\Psi\rangle-\langle\Psi|\hat
{n}_{a}|\Psi\rangle\langle\Psi|\hat{n}_{b}|\Psi\rangle$.

Using Mandel $Q$-parameter $Q_{i}=(\Delta^{2}\hat{n}_{i}-\left\langle \hat
{n}_{i}\right\rangle )/\left\langle \hat{n}_{i}\right\rangle $ $(i=a$, $b)$
\cite{Mandel} to describe the intramode correlations and $J=Cov[\hat{n}%
_{a},\hat{n}_{b}]/\Delta\hat{n}_{a}\Delta\hat{n}_{b}$ ($-1\leq J\leq1$)
\cite{Gerrybook} to describe the intermode correlations, Eq.~(\ref{Fisher})
can be written as%
\begin{align}
\mathcal{F}  &  =\left\langle \hat{n}_{a}\right\rangle (Q_{a}+1)+\left\langle
\hat{n}_{b}\right\rangle (Q_{b}+1)\nonumber\\
&  +2\sqrt{\left\langle \hat{n}_{a}\right\rangle \left\langle \hat{n}%
_{b}\right\rangle (Q_{a}+1)(Q_{b}+1)}J, \label{Fisher2}%
\end{align}
where $\left\langle \hat{n}_{i}\right\rangle =\langle\Psi|\hat{n}_{i}%
|\Psi\rangle$ $(i=a$, $b)$. The QFI $\mathcal{F}$ is composed of the photon
statistics in each arm and the correlation between two arms. The transform of
the operators by the first NBS of SU(1,1) interferometer is $\hat{a}%
_{out}=u\hat{a}_{in}+v\hat{b}_{in}^{\dagger}$, $\hat{b}_{out}=u\hat{b}%
_{in}+v\hat{a}_{in}^{\dagger}$, where $u=\cosh g_{1}$ and $v=\sinh
g_{1}e^{i\theta_{1}}$. $g_{1}$ describes the strength in the process of first
NBS, and $\theta_{1}$ is controlled by the phase of the pump field as shown in
Fig.~\ref{fig1}. For convenience, we use $g$ to replace the $g_{1}$ because we
only consider the first NBS in this proposal. Compared with the traditional
MZI, the phase sensitivity of SU(1,1) interferometer can be improved due to
the amplification process of the NBS. For vacuum state input, $\left\langle
\hat{n}_{a}\right\rangle =\left\langle \hat{n}_{b}\right\rangle =\left\vert
v\right\vert ^{2}=\bar{N}_{\text{inside}}/2$, $\Delta\hat{n}_{a}=\Delta\hat
{n}_{b}$, the Eq.~(\ref{Fisher2}) can be reduced as $\mathcal{F}=2\bar
{N}_{\text{inside}}(Q+1)$, where $J=1$. For SU(1,1) interferometers, any input
states passing through the first NBS the correlation between mode $\hat{a}$
and mode $\hat{b}$ will be generated, which leads to $J>0$ \cite{Yuan16} (see
Eq.~(32) in it). For an SU(1,1) interferometer, whatever the input state
chosen, the lower and upper bounds of QFI are given by: \begin{widetext}%
\begin{equation}
\left[  \left\langle \hat{n}_{a}\right\rangle (Q_{a}+1)+\left\langle \hat
{n}_{b}\right\rangle (Q_{b}+1)\right]  <\mathcal{F}\leq\left(  \sqrt
{\left\langle \hat{n}_{a}\right\rangle (Q_{a}+1)}+\sqrt{\left\langle \hat
{n}_{b}\right\rangle (Q_{b}+1)}\right)  ^{2}.
\end{equation}
\end{widetext}The above inequality shows that the metrological advantage of
nonclassical light is primarily the photon statistics-Mandel parameters
$Q_{a}$ and $Q_{b}$. The intermode correlations can contribute at most a
factor of $2$ improvement in the QFI, hence mode entanglement is not a
necessary resource for quantum metrology which is the same as that of MZI
\cite{Sahota16}. Therefore, we need to search and study quantum states where
photon statistics within each arm of the interferometer should be
super-Poissonian ($Q>0$) and with high intramode correlations simultaneously.
For MZI, beams splittering and recombination are linear process and no
amplification. Under the condition of path symmetry assumption $\left\langle
\hat{n}_{a}\right\rangle =\left\langle \hat{n}_{b}\right\rangle =\bar{n}/2$
and $\left\langle \hat{n}_{a}^{2}\right\rangle =\left\langle \hat{n}_{b}%
^{2}\right\rangle $, the QFI $\mathcal{F}$ in terms of intramode and intermode
correlations was studied by Sahota and Quesada \cite{Sahota} and was given as
follows:
\begin{equation}
\mathcal{F}=\bar{n}(Q+1)(1-J).
\end{equation}
Note that $-J$ is for MZI due to the differences between $\hat{J}_{z}$ and
$\hat{K}_{z}$.

Next, we study a coherent light combined with a photon-subtracted squeezed
vacuum light as input in an SU(1,1) interferometer, due to the
super-Poissonian statistical properties of photon-subtracted squeezed vacuum
state \cite{Gerry}, and compare the phase sensitivities with the HL.

\subsection{QFI of coherent mixed photon subtracted squeezed vacuum states}

\begin{figure}[ptb]
\includegraphics[scale=0.45,angle=0]{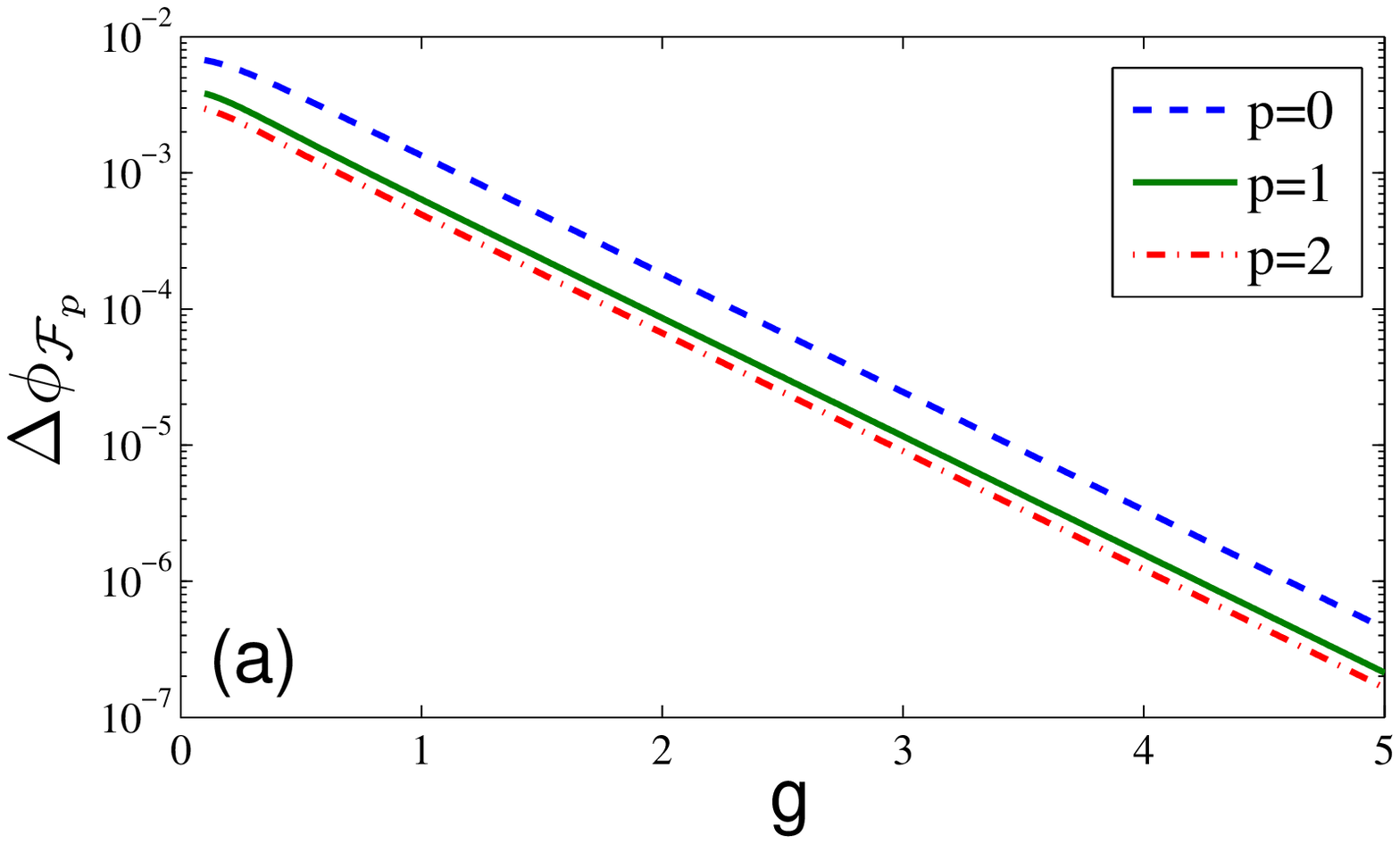}
\includegraphics[scale=0.45,angle=0]{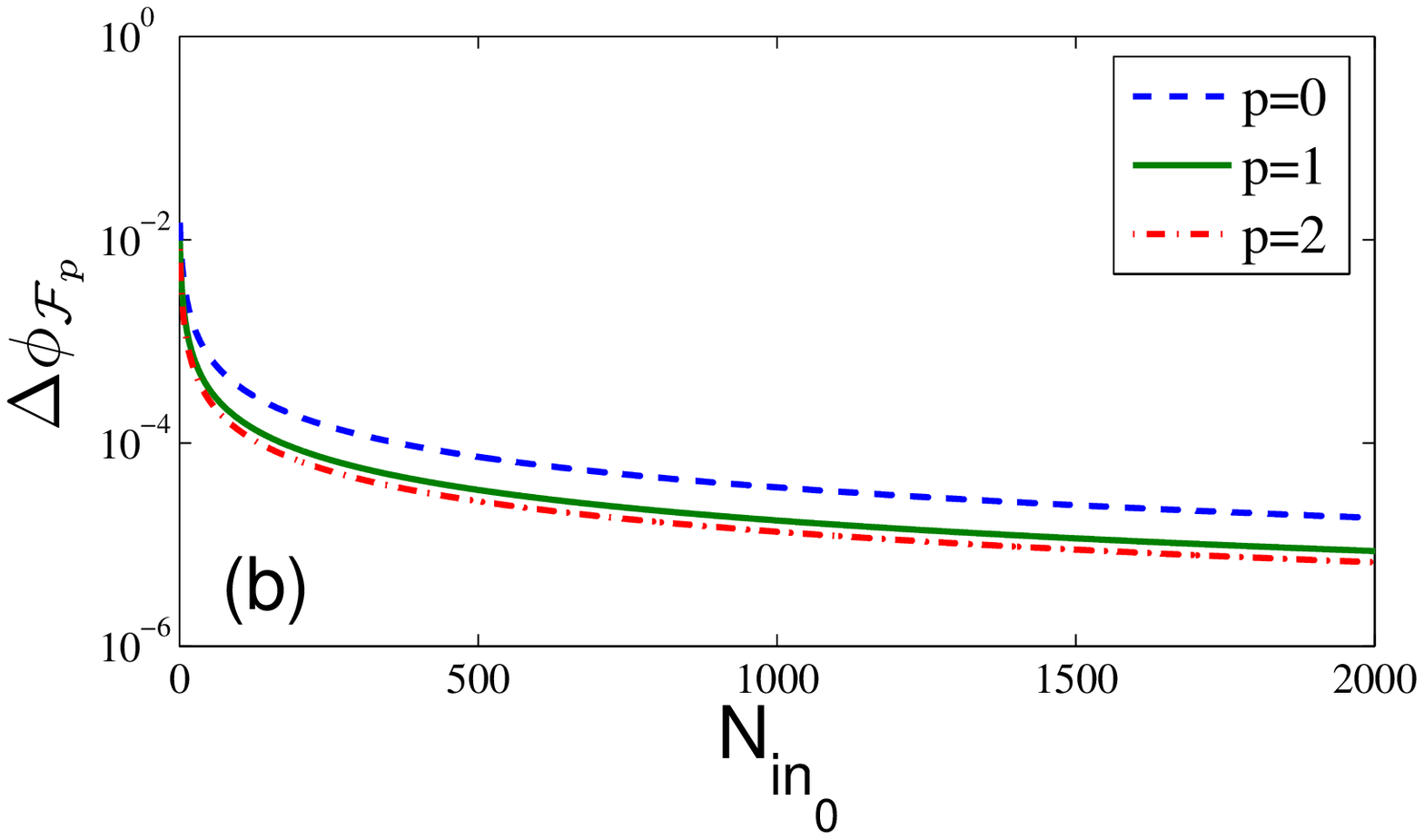}
\includegraphics[scale=0.45,angle=0]{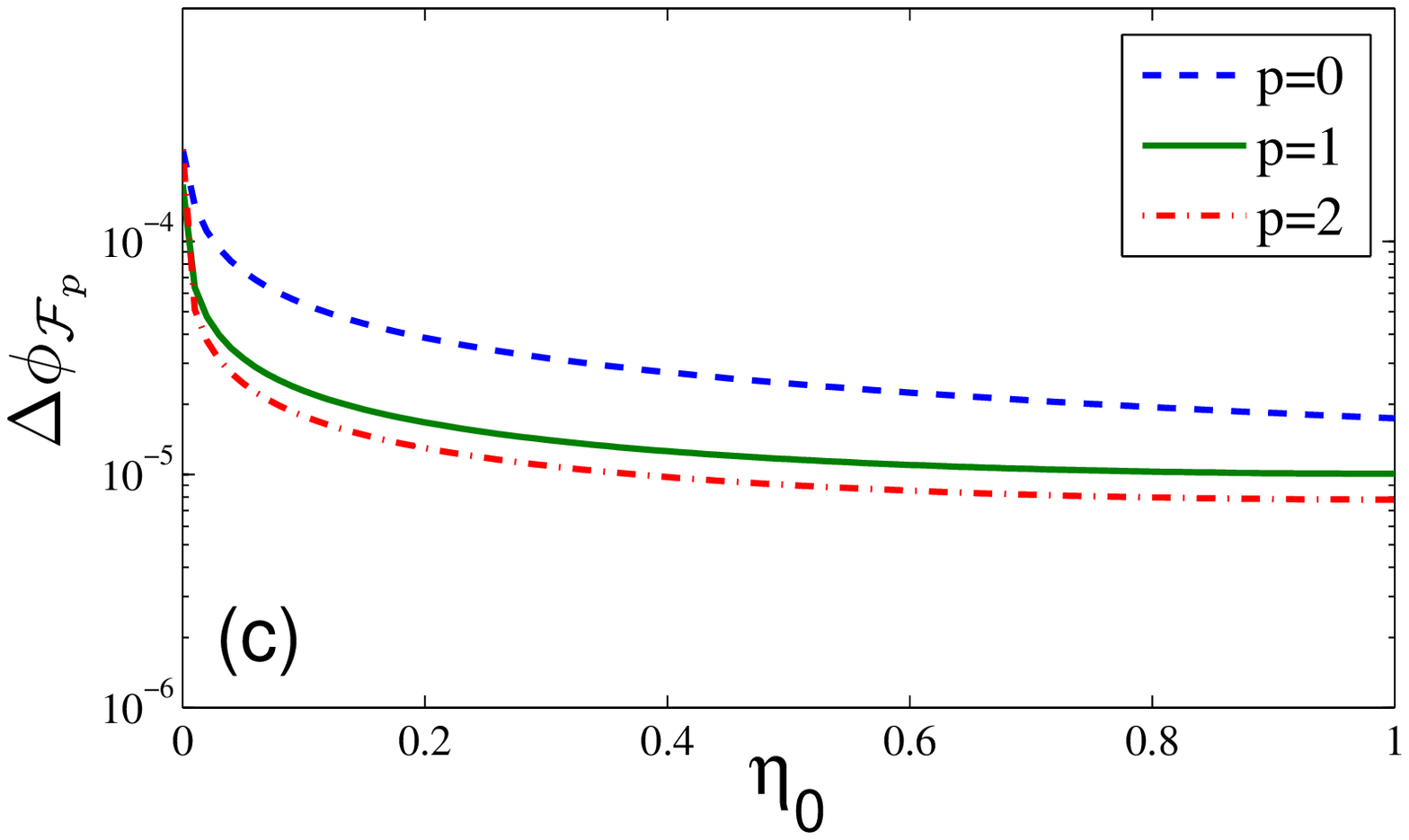} \caption{(Color online) The
phase sensitivities $\Delta\phi_{\mathcal{F}_{p}}$ of the SU(1,1)
interferometer (a) versus $g$ with $N_{\text{in}_{0}}=200$; (b) versus
$N_{\text{in}_{0}}$ with $g=3$ where $\eta_{0}=0.5$; (c) versus $\eta_{0}$
with $g=3$ and $N_{\text{in}_{0}}=200$.}%
\label{fig2}%
\end{figure}

As shown in Fig.~\ref{fig1}, a BS with low reflectance and a single-photon
resolution photo-detector are used to subtract $p$ photons from input squeezed
vacuum state. The reflected photons that are detected herald a $p$-photon
subtracted squeezed-vacuum state generation \cite{Biswas}%
\begin{equation}
|p,\varsigma\rangle_{b}\sim\hat{b}^{p}|0,\varsigma\rangle_{b},
\end{equation}
where $|0,\varsigma\rangle_{b}=\hat{S}_{b}(r)|0\rangle_{b}$ is a single-mode
squeezed vacuum state in the $b$-mode, and $\hat{S}_{b}(r)=\exp[(-\varsigma
\hat{b}^{\dag2}+\varsigma^{\ast}\hat{b}^{2})/2]$ with $\varsigma=r\exp
(i\theta_{\varsigma})$ is the single-mode squeezing parameter. We consider a
coherent light combined with a $p$-photon subtracted squeezed vacuum light as
input, i.e., $|\Psi_{\text{in}}\rangle=|\alpha\rangle_{a}\otimes
|p,\varsigma\rangle_{b}$, where $\alpha=\left\vert \alpha\right\vert
e^{i\theta_{\alpha}}$. Here, we use this state as the probe state in phase
sensitivity measurement. We firstly derive the QFI $\mathcal{F}_{p}$
$(p=0,1,2)$.

For $p=0$, the input state $|\Psi_{\text{in}}\rangle$ is the coherent state
$\otimes$ squeezed vacuum state, which has been studied by some of us only
with the method of error propagation using the homodyne detection \cite{Li14}
and parity detection \cite{Li16}. For $p=1$ and $p=2$, the $p$-photon
subtracted squeezed vacuum states $|p,\varsigma\rangle_{b}$ are respectively
given by
\begin{equation}
|1,\varsigma\rangle=\frac{1}{\sqrt{\bar{n}_{0}}}\hat{b}|0,\varsigma\rangle
_{b},
\end{equation}
and
\begin{equation}
|2,\varsigma\rangle=\frac{1}{\sqrt{\bar{n}_{0}\bar{n}_{1}}}\hat{b}%
^{2}|0,\varsigma\rangle_{b},
\end{equation}
where $\bar{n}_{0}=\sinh^{2}r$ and $\bar{n}_{1}=\sinh^{2}r+\cosh(2r)$. Using
the coherent state combined with a $p$-photon subtracted squeezed vacuum state
as input ($p=0$, $1$, $2$), when $\theta_{\varsigma}+2\theta_{\alpha}%
-2\theta_{1}=\pi$, according to Eq.~(\ref{Fisher}) the maximal QFIs of them
are given by%
\begin{align}
\mathcal{F}_{0}  &  =\cosh^{2}(2g)\left[  \frac{1}{2}\sinh^{2}(2r)+\left\vert
\alpha\right\vert ^{2}\right]  +\sinh^{2}(2g)\nonumber\\
&  \times\left[  \left\vert \alpha\right\vert ^{2}e^{2r}+\bar{n}_{0}+1\right]
,
\end{align}%
\begin{align}
\mathcal{F}_{1}  &  =\cosh^{2}(2g)\left[  \frac{3}{2}\sinh^{2}(2r)+\left\vert
\alpha\right\vert ^{2}\right]  +\sinh^{2}(2g)\nonumber\\
&  \times\lbrack3\left\vert \alpha\right\vert ^{2}e^{2r}+\bar{n}_{1}+1],
\end{align}%
\begin{align}
\mathcal{F}_{2}  &  =\cosh^{2}(2g)[\frac{3}{2}\sinh^{2}(2r)\frac{5\sinh
^{2}r(\bar{n}_{1}+1)+3}{\bar{n}_{1}^{2}}\nonumber\\
&  +\left\vert \alpha\right\vert ^{2}]+\sinh^{2}(2g)\{\left\vert
\alpha\right\vert ^{2}[3\sinh(2r)\frac{5\sinh^{2}r+1}{\bar{n}_{1}}\nonumber\\
&  +2\bar{n}_{2}+1]+\bar{n}_{2}+1\},
\end{align}
where $\bar{n}_{2}=3\sinh^{2}r(5\sinh^{2}r+3)/\bar{n}_{1}$. Here $\bar{n}_{p}$
are the mean photon number of $p$-photon subtracted squeezed vacuum states
($p=0$, $1$, $2$). $\mathcal{F}_{0}$ has been studied by some of us
\cite{Gong16}, and here it is used for comparison with $\mathcal{F}_{1}$\ and
$\mathcal{F}_{2}$.

Next, we give the QCRB according to the QFI, which is given by%
\begin{equation}
\Delta\phi_{\mathcal{F}}=\frac{1}{\sqrt{m\mathcal{F}_{p}}}\text{
\ \ }(p=0\text{, }1\text{, }2), \label{QCRB}%
\end{equation}
where $m$ is the number of independent repeats of the experiment, and
subscript $p$ denotes subtracting $p$ photons from the squeezed vacuum state.
Using Eq.~(\ref{QCRB}) we obtain the phase sensitivities $\Delta
\phi_{\mathcal{F}_{p}}$, which increase with increasing $g$ and $N_{\text{in}%
_{0}}$ as shown in Figs. \ref{fig2} (a) and (b). For fixed $N_{\text{in}_{0}}$
and $g$, the phase sensitivities $\Delta\phi_{\mathcal{F}_{p}}$ also improve
with increasing the photon subtractions $p$ ($p=0$, $1$, $2$). For given a
fixed input mean number of photons $N_{\text{in}_{0}}$, the subtraction of
photons from the squeezed vacuum state increases the corresponding sensitivity
in the phase-shift measurement. Therefore, for the coherent state $\otimes$
the squeezed vacuum state input, given a fixed input mean number of photons
$N_{\text{in}_{0}}$, the phase sensitives $\Delta\phi_{\mathcal{F}}$ can be
improved by subtracting the photons from the squeezed vacuum state, which
seems to conflict with the result of Lang and Caves \cite{Lang} and we explain
it in Sec. IV.

To describe the effect of unbalanced input states on the QFI, we introduce a
parameter $\eta_{p}$ which is defined by
\begin{align}
\eta_{p}  &  \equiv\frac{\text{mean photon number of }b\text{ mode input}%
}{\text{total mean photon number of input}}\nonumber\\
&  =\bar{n}_{p}/N_{\text{in}_{p}}\text{ }(p=0,1,2),
\end{align}
where the total mean photon numbers into the first NBS are written as
$N_{\text{in}_{p}}=\left\vert \alpha\right\vert ^{2}+\bar{n}_{p}$ ($p=0$, $1$,
$2$). For $p=0$ the squeezed vacuum state with coherent state input, when
$\eta_{0}=0$, the input state is a coherent state $|\alpha\rangle$, and when
$\eta_{0}=1$, it is a squeezed vacuum state input. For a given fixed
$N_{\text{in}_{0}}$, in order to find the dominant component of $r$ and
$\left\vert \alpha\right\vert $, the phase sensitivities $\Delta
\phi_{\mathcal{F}}$ as a function of the squeezing fraction $\eta_{0}$ is
shown in Fig. \ref{fig2}(c). We find that the optimal squeezing fraction
$\eta_{0}$\ is $1$ for a given fixed $N_{\text{in}_{0}}$. That is for a given
fixed $N_{\text{in}_{0}}$, only with $p$-photon subtracted squeezed vacuum
light as input and without the coherent state, the phase sensitivities are the
highest $(p=0,1,2)$.

\section{Heisenberg limit}

\begin{figure}[ptb]
\includegraphics[scale=0.42,angle=0]{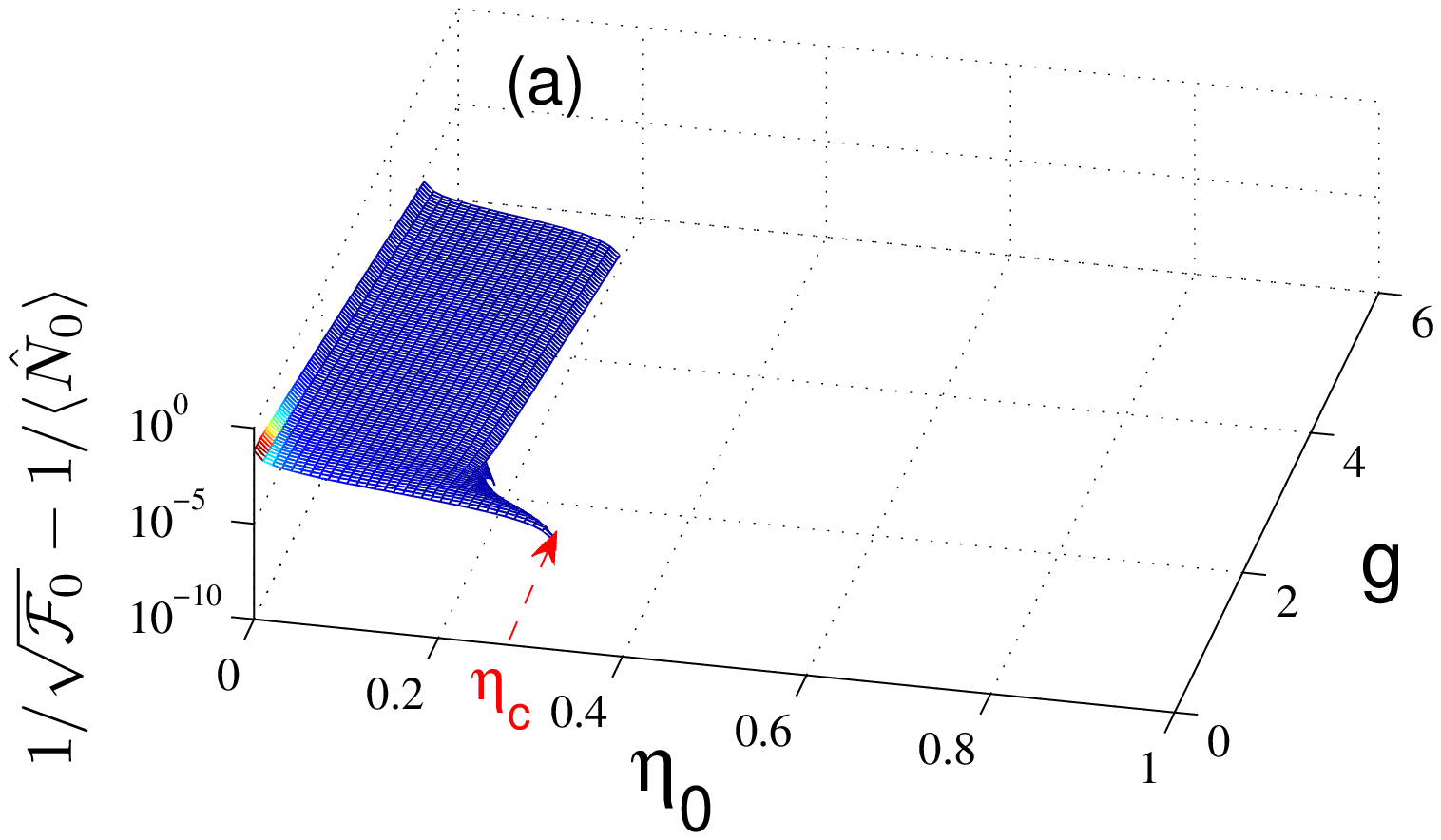}
\includegraphics[scale=0.42,angle=0]{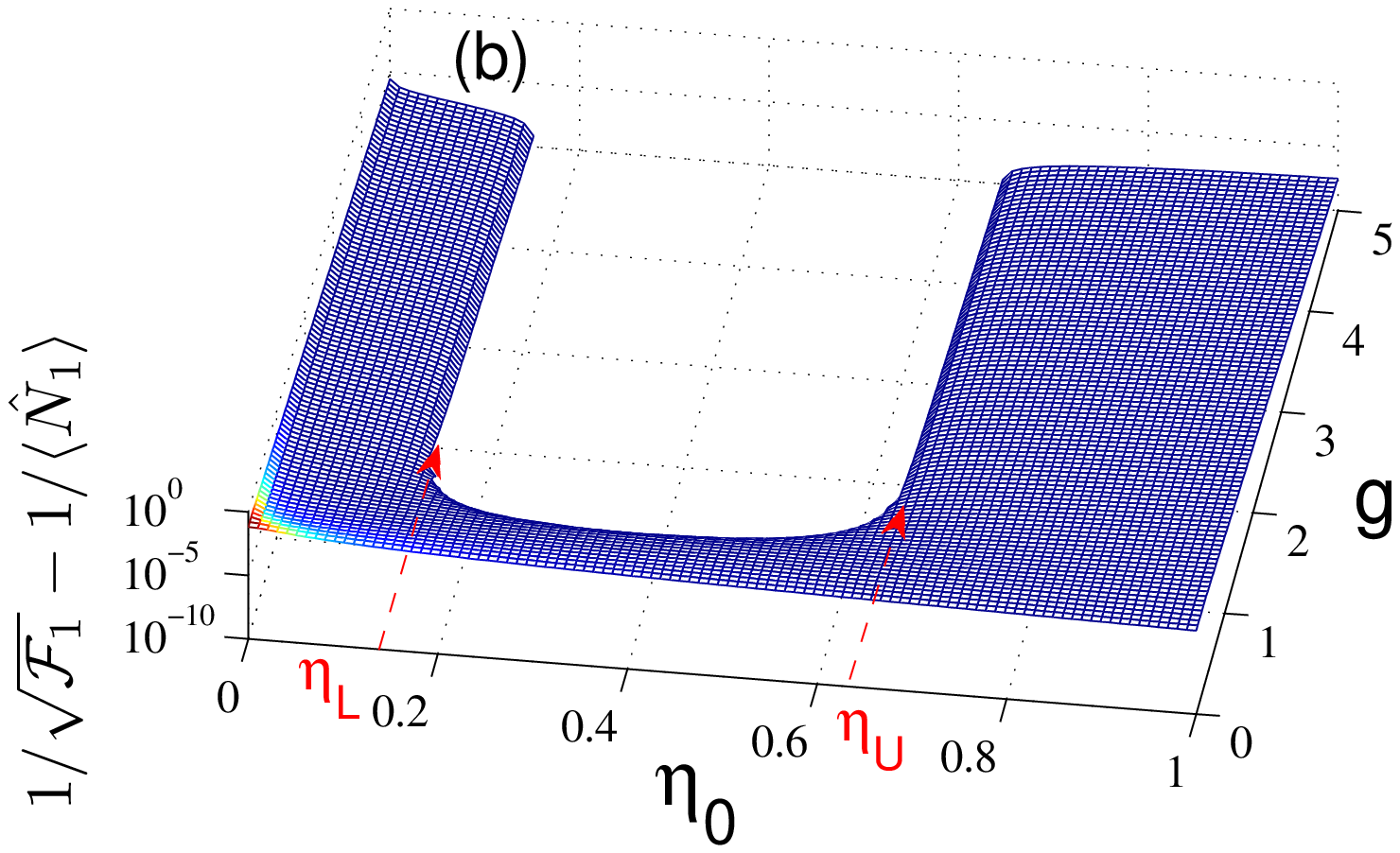}
\includegraphics[scale=0.42,angle=0]{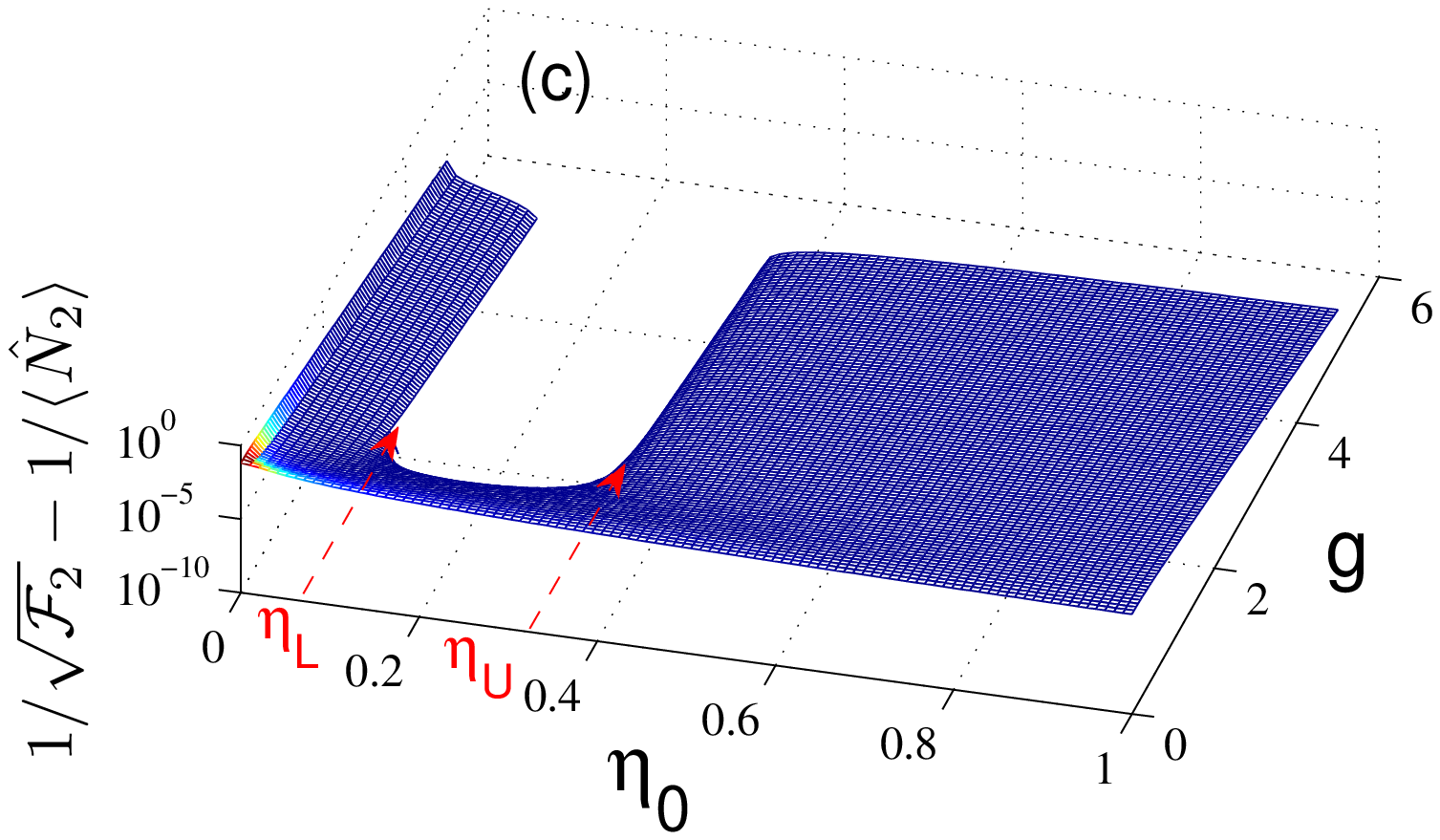} \caption{(Color online) Under
the condition of small-$m$ case ($m=1$), the difference between the phase
sensitivities $\Delta\phi_{\mathcal{F}_{p}}$ $(=1/\sqrt{\mathcal{F}_{p}})$ and
HL $\Delta\phi_{\mathrm{HL}_{p}}$ $(=1/\langle\hat{N}_{p}\rangle)$ as a
function of $\eta_{0}$ and $g$ with $N_{\mathrm{in}_{0}}=200$ for (a) $p=0$;
(b) $p=1$; (c) $p=2$. In the blank areas of $\eta_{\text{0}}>\eta_{c}$ and
$\eta_{\text{L}}<\eta_{0}<\eta_{\text{U}}$, the phase sensitivities
$\Delta\phi_{\mathcal{F}}$ can beat the the scaling $1/\langle\hat{N}\rangle$
under a certain condition.}%
\label{fig3}%
\end{figure}

The HL is ultimate scaling of the phase sensitivity imposed by quantum
mechanics. For the number of particles in the probe state is fixed and is
equal to $N$ (without number fluctuations), the fundamental HL is given by
\cite{pezze2015}
\begin{equation}
\Delta\phi_{\text{HL}}=\frac{1}{\sqrt{m}N},\label{HLold}%
\end{equation}
where the factor $m$ is the number of independent measurements repeated with
identical copies of the probe state. In the SU(2) scheme, the optimal input
state is the NOON state and the sensitivity is $1/N$ \cite{NOON,Dowling}.
However, for a fixed mean photon number (with number fluctuations) the form of
the HL has been questioned \cite{Sharpio,Pasquale}. Considering the photon
number fluctuations, Hofmann suggested the form of HL is $1/\langle\hat{N}%
^{2}\rangle^{1/2}$, where indicates averaging over the squared photon numbers
\cite{hofmann}. But the phase sensitivity can be arbitrary high for large
fluctuations. For separable state and with unbiased phase estimators,
Pezz\`{e} \textit{et al.} pointed out that the HL for two-mode interferometers
should add extra constraints and is given by \cite{pezze2015,hyllus}
\begin{equation}
\Delta\phi_{\text{HL}}=\max\left[  \frac{1}{\sqrt{m\langle\hat{N}^{2}\rangle}%
},\frac{1}{m\langle\hat{N}\rangle}\right]  ,\text{ for }\langle\Delta\hat
{N}\rangle>0.\label{HLnew}%
\end{equation}
Eq.~(\ref{HLnew}) cannot be obtained from Eq.~(\ref{HLold}) by simply
replacing $N$ with $\langle\hat{N}\rangle$. But Eq.~(\ref{HLnew}) can reduce
to Eq.~(\ref{HLold}) when number fluctuations vanish, i.e., $\langle\hat{N}%
^{2}\rangle=\langle\hat{N}\rangle^{2}=N^{2}$.

For the SU (1,1) interferometer, $\langle\hat{N}\rangle$ $(\equiv\langle
\Psi|\hat{n}_{a}+\hat{n}_{b}|\Psi\rangle)$ is the total number of photons
inside the interferometer~\cite{Marino}, not the input one as the traditional
MZI. This is due to phase sensing by the total photons of the two modes inside
the interferometer. The mean photon numbers inside the interferometer for
$p=0$, $1$ and $2$ are respectively given by%
\begin{equation}
\langle\hat{N}_{p}\rangle=\cosh(2g)N_{\text{in}_{p}}+2\sinh^{2}(g),\text{
}(p=0,1,2).
\end{equation}
Similarly, one can obtain the squared photon numbers inside the SU(1,1)
interferometer, which are given as following:
\begin{align}
\langle\hat{N}_{0}^{2}\rangle &  =\left(  \left\vert \alpha\right\vert
^{4}+3\bar{n}_{0}^{2}\right)  \cosh^{2}(2g)+4\left(  N_{\text{in}_{0}%
}+1\right)  \sinh^{4}(g)\nonumber\\
&  +[\left\vert \alpha\right\vert ^{2}\cosh(2r)+2\bar{n}_{0}]\cosh
(4g)+\sinh^{2}(2g)\nonumber\\
&  \times\lbrack\left\vert \alpha\right\vert ^{2}(\sinh(2r)+1)+1],
\end{align}
$\ \ \ \ \ \ \ $%
\begin{align}
\langle\hat{N}_{1}^{2}\rangle &  =\cosh(4g)[\left\vert \alpha\right\vert
^{2}(6\bar{n}_{0}+1)+2N_{\text{in}_{1}}-1]+\cosh^{2}(2g)\nonumber\\
&  \times(15\bar{n}_{0}^{2}+\left\vert \alpha\right\vert ^{4}+6\bar{n}%
_{0})+4\sinh^{4}g(N_{\text{in}_{1}}+1)\nonumber\\
&  +\sinh^{2}(2g)[\left\vert \alpha\right\vert ^{2}+2+3\left\vert
\alpha\right\vert ^{2}\sinh(2r)],
\end{align}

\begin{align}
&  \langle\hat{N}_{2}^{2}\rangle=\cosh(4g)[2\bar{n}_{2}+\left\vert
\alpha\right\vert ^{2}(2\bar{n}_{2}+1)]+4\sinh^{4}(g)(1\nonumber\\
&  +N_{\text{in}_{2}})+\cosh^{2}(2g)\left[  35\bar{n}_{0}^{2}+\frac{1}%
{3\bar{n}_{0}+1}40\bar{n}_{0}^{2}+\left\vert \alpha\right\vert ^{4}\right]
\nonumber\\
&  +\sinh^{2}(2g)\left[  \left\vert \alpha\right\vert ^{2}+3\frac{5\bar{n}%
_{0}+1}{3\bar{n}_{0}+1}\left\vert \alpha\right\vert ^{2}\sinh2r+1\right]  ,
\end{align}
where $\theta_{\varsigma}+\theta_{\alpha}-\theta_{1}=\pi$.

Next, we compare the QCRB with the HL from Eq.~(\ref{HLnew}). According to the
number of measurements $m$, we study two cases: (i) small-$m$ case; (ii)
large-$m$ case. As discussed in Ref.~\cite{pezze2015}, in the limit of
small-$m$ the HL of Eq.~(\ref{HLnew}) is found to be
\begin{equation}
\Delta\phi_{\text{HL}p}=\frac{1}{m\langle\hat{N}_{p}\rangle}\text{
\ }(p=0\text{, }1\text{, }2). \label{HL2}%
\end{equation}
In this situation, we consider $m=1$, and the HLs are given by%
\begin{equation}
\Delta\phi_{\text{HL}p}=\frac{1}{\langle\hat{N}_{p}\rangle}\text{
\ \ }(p=0\text{, }1\text{, }2).
\end{equation}
For a given input number, the difference between the phase sensitivity
$\Delta\phi_{\mathcal{F}_{p}}$ and the HL $\Delta\phi_{\text{HL}p}$ ($p=0$,
$1$, $2$) as a function of $\eta_{0}$ and $g$ is shown in Fig. \ref{fig3}.
When $\eta_{\text{0}}>\eta_{c}$ and $\eta_{\text{L}}<\eta_{0}<\eta_{\text{U}}%
$, the phase sensitivities $\Delta\phi_{\mathcal{F}_{p}}$ can beat the the
scaling $1/\langle\hat{N}_{p}\rangle$ ($\Delta\phi_{\mathcal{F}_{p}}-$
$\Delta\phi_{\text{HL}p}<0$, $p=0$, $1$, $2$) within certain parameters, as
shown in the blank areas of Fig. \ref{fig3}. For $p=$ $1\ $and $2$ cases, with
a given $N_{\text{in}_{0}}$\ and $g$, although the phase sensitivities
$\Delta\phi_{\mathcal{F}}$ are optimal under the condition of $\eta_{0}=1$
shown in Fig. \ref{fig2}(c), the areas of $\Delta\phi_{\mathcal{F}_{p}}-$
$\Delta\phi_{\text{HL}p}<0$ is $\eta_{\text{L}}<\eta_{0}<\eta_{\text{U}}$ as
shown in Fig. \ref{fig3}(b) and (c). The reason is that both the phase
sensitivity $\Delta\phi_{\mathcal{F}}$ and the HL $\Delta\phi_{\text{HL}}$
enhance with increase of $\eta_{0}$, but the increasing rate of $\Delta
\phi_{\text{HL}}$ is higher.

\begin{figure}[ptb]
\includegraphics[scale=0.45,angle=0]{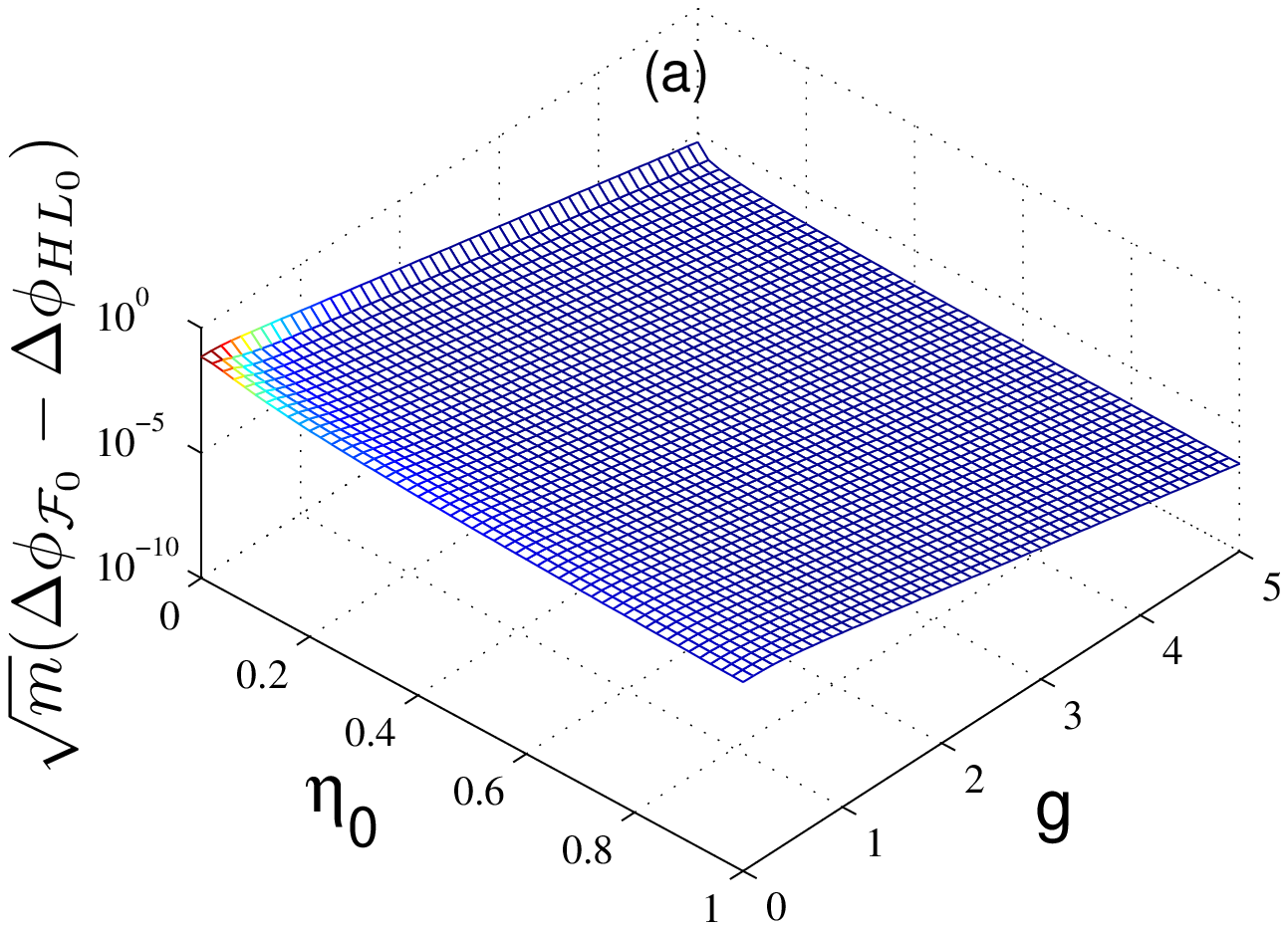}
\includegraphics[scale=0.45,angle=0]{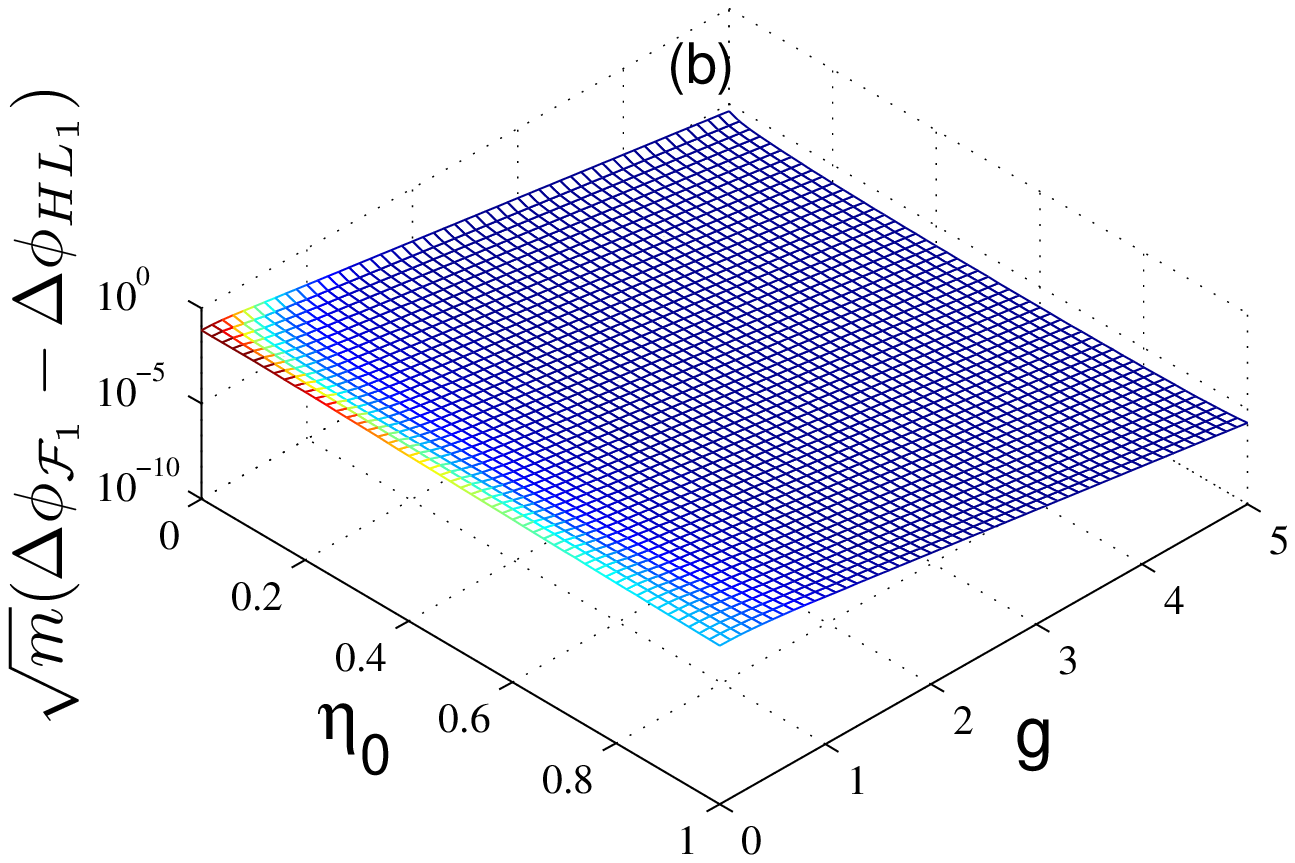}
\includegraphics[scale=0.45,angle=0]{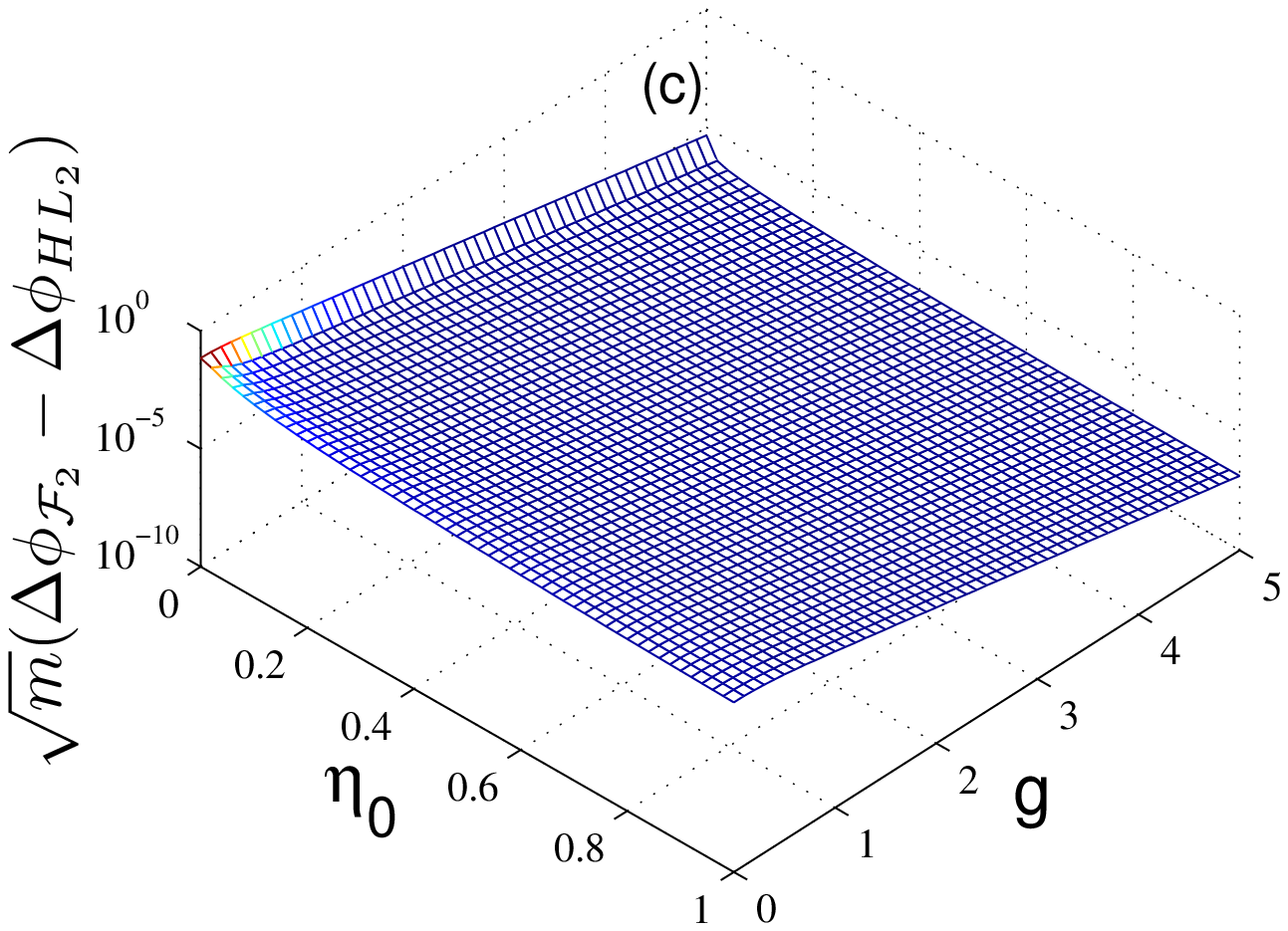} \caption{(Color online) Under
the condition of large-$m$ case, the difference between the phase
sensitivities $\Delta\phi_{\mathcal{F}_{p}}$ $(=1/\sqrt{m\mathcal{F}_{p}})$
and HL $\Delta\phi_{\mathrm{HL}_{p}}$ $( =1/\sqrt{m\langle\hat{N}_{p}%
^{2}\rangle}) $ as a function of $\eta_{0}$ and $g$ with $N_{\mathrm{in}_{0}%
}=200$ for (a) $p=0$; (b) $p=1$; (c) $p=2$.}%
\label{fig4}%
\end{figure}

Now we will study the case of large-$m$ limit. As shown in
Ref.~\cite{pezze2015}, the HL of Eq.~(\ref{HLnew}) in the large-$m$ limit is
given by
\begin{equation}
\Delta\phi_{\text{HL}p}=\frac{1}{\sqrt{m\langle\hat{N}_{p}^{2}\rangle}}\text{
\ \ }(p=0\text{, }1\text{, }2). \label{HL}%
\end{equation}
We compare the corresponding HL of Eq.~(\ref{HL}) with the phase sensitivity
by QCRB in Figs.~\ref{fig4}(a)-(c) under various input situations, and obtain
that the HL of large-$m$ limit case cannot be beaten.

The comparison between two different results shown in Figs.~\ref{fig3} and
\ref{fig4} deserves a discussion. In the small-$m$ limit ($m=1$ particularly),
for the $p$-photon subtracted squeezed vacuum states input the phase
sensitivity QCRB $\Delta\phi_{\mathcal{F}_{p}}=1/\sqrt{\mathcal{F}_{p}}$ can
beat $1/\langle\hat{N}_{p}\rangle$. While in the large-$m$ limit, this
ultimate quantum limit of SU(1,1) interferometer is $\Delta\phi_{\text{HL}%
p}=1/\sqrt{m\langle\hat{N}_{p}^{2}\rangle}$, and it cannot be beaten. As we
all know, the HL is as the maximum sensitivity achievable, optimized over all
possible estimators, measurements and probe states. However, in presence of
large fluctuations in the number of probes, the violation of $1/\langle\hat
{N}\rangle$ scaling of the optimal accuracy is possible
\cite{Sharpio,Pasquale}, and the definition of the HL should take into account
the amount of fluctuations \cite{hofmann,pezze2015}. Hence, this HL
$\Delta\phi_{\text{HL}p}=1/\sqrt{m\langle\hat{N}_{p}^{2}\rangle}$ considering
the number fluctuation effect could be the ultimate phase sensitivity
limit~\cite{hofmann,hyllus,pezze2015}.

\section{Discussion}

\begin{figure}[ptb]
\includegraphics[scale=0.45,angle=0]{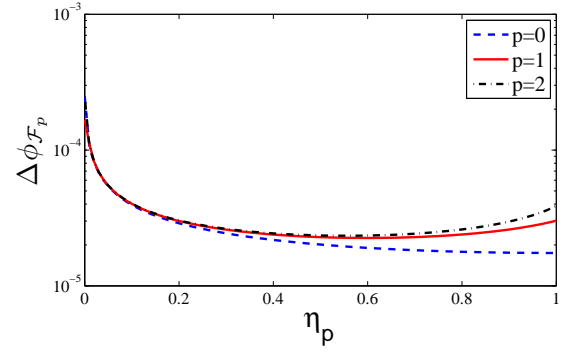}\caption{(Color online) The
phase sensitivities $\Delta\phi_{\mathcal{F}_{p}}$ as a function of $\eta_{p}$
$(p=0,1,2)$ with $N_{\text{in}_{0}}=N_{\text{in}_{1}}=N_{\text{in}_{2}}=200$
and $g=3$. }%
\label{fig5}%
\end{figure}

For the coherent state $\otimes$ the squeezed vacuum state input, given a
fixed input mean number of photons $N_{\text{in}_{0}}$, the phase sensitives
$\Delta\phi_{\mathcal{F}}$ can be improved by subtracting the photons from the
squeezed vacuum state, as shown in Fig. \ref{fig2}. That is within the
constraint on the average photon of a squeezed vacuum state, it is still
possible to attain higher sensitivity via photon subtraction. It seems to
conflict with the result of proposal presented by Lang and Caves \cite{Lang}.
They considered that an interferometer powered by laser light (a coherent
state) into one input port and ask the following question: what is the best
state to inject into the second input port, given a constraint on the mean
number of photons this state can carry, in order to optimize the
interferometer's phase sensitivity? Their answer is squeezed vacuum. In fact,
they don't conflict and the reason is given as follows.

With $N_{\text{in}_{p}}$ and $\eta_{p}$, the maximal QFIs $\mathcal{F}_{p}$
($p=0$, $1$, $2$) are rewritten as
\begin{align}
\mathcal{F}_{0}  &  =\cosh^{2}(2g)[\eta_{0}N_{\text{in}_{0}}(1+2\eta
_{0}N_{\text{in}_{0}})+N_{\text{in}_{0}}]+\sinh^{2}(2g)\nonumber\\
&  \times\lbrack2\eta_{0}(1-\eta_{0})(N_{\text{in}_{0}})^{2}+2\sqrt{\eta
_{0}N_{\text{in}_{0}}(\eta_{0}N_{\text{in}_{0}}+1)}\nonumber\\
&  \times(1-\eta_{0})N_{\text{in}_{0}}+N_{\text{in}_{0}}+1],
\end{align}

\begin{align}
\mathcal{F}_{1}  &  =\cosh^{2}(2g)[\frac{2}{3}(\eta_{1}N_{\text{in}_{1}%
}-1)(\eta_{1}N_{\text{in}_{1}}+2)+(1-\eta_{1})\nonumber\\
&  \times N_{\text{in}_{1}}]+\sinh^{2}(2g)[2\eta_{1}(N_{\text{in}_{1}}%
)^{2}(1-\eta_{1})+N_{\text{in}_{1}}+1\nonumber\\
&  +2N_{\text{in}_{1}}(1-\eta_{1})\sqrt{(\eta_{1}N_{\text{in}_{1}}-1)(\eta
_{1}N_{\text{in}_{1}}+2)}],
\end{align}%
\begin{align}
\mathcal{F}_{2}  &  =\cosh^{2}(2g)[6S(S+1)\frac{5S(3S+2)+3}{(3S+1)^{2}%
}+(1-\eta_{2})\nonumber\\
&  \times N_{\text{in}_{2}}]+\sinh^{2}(2g)\{N_{\text{in}_{2}}(1-\eta
_{2})[6\sqrt{S(S+1)}\nonumber\\
&  \times\frac{5S+1}{3S+1}+2\eta_{2}N_{\text{in}_{2}}+1]+\eta_{2}%
N_{\text{in}_{2}}+1\},
\end{align}
where
\begin{equation}
S=\frac{\eta_{2}N_{\text{in}_{2}}-3+\sqrt{\left(  \eta_{2}N_{\text{in}_{2}%
}\right)  ^{2}+\frac{2}{3}\eta_{2}N_{\text{in}_{2}}+9}}{10}.
\end{equation}
Let $N_{\text{in}_{0}}=N_{\text{in}_{1}}=N_{\text{in}_{2}}$ and for a fixed
$g$, the phase sensitivities $\Delta\phi_{\mathcal{F}_{p}}$ as a function of
the squeezing fraction $\eta_{p}$ is shown in Fig. \ref{fig5}, and we obtain
$\Delta\phi_{\mathcal{F}_{0}}<\Delta\phi_{\mathcal{F}_{1}}<\Delta
\phi_{\mathcal{F}_{2}}$. which agrees with the result given by Lang and Caves
\cite{Lang}. However, the reflected photons that are detected herald a
photon-subtracted squeezed-vacuum state generation, the sensitivity of
phase-shift estimation can also be improved. The generation process of
photon-subtracted squeezed-vacuum state need to use additional resources and
is probability.

\section{Conclusions}

We have studied the phase sensitivities of an SU(1,1) interferometer with a
coherent state in one input port and a photon-subtracted squeezed vacuum state
in the other input port using QCRB. The subtraction of photons from the
squeezed vacuum state not only increases the average photon number for the
fixed squeezed parameter $r$, but also increases the corresponding sensitivity
in the phase-shift measurement. The HL considering the number fluctuation
effect cannot be beaten and may be the ultimate phase limit. The enhancement
of phase sensitivity with nonclassical light is dominant from the intramode
correlations, hence the mode entanglement is not a critical resource for
quantum metrology. Using separable states have many advantages over entangled
states including more flexibility in the distribution of resources,
comparatively easier state preparation.

\section*{Acknowledgements}

This work was supported by the National Key Research Program of China under
Grant number Grant No.~2016YFA0302000 and the National Natural Science
Foundation of China under Grant Nos.~11474095, ~11234003, and~11129402, and
the Fundamental Research Funds for the Central Universities.



\begin{thebibliography}{99}
\bibitem {TMS}Two Hermitian operators $\hat{K}_{x}$\ and $\hat{K}_{y}$
generate the two-mode squeezing operator $\hat{S}_{ab}(\xi)=\exp(-\xi\hat
{a}^{\dag}\hat{b}^{\dag}+\xi^{\ast}\hat{a}\hat{b})=\exp\left[  -\frac{i}%
{\hbar}(2\xi_{2}\hat{K}_{x}+2\xi_{1}\hat{K}_{y})\right]  $, where $\xi=\xi
_{1}+i\xi_{2}=re^{i\theta_{\xi}}$ is the squeezed parameter of the nonlinear
beam splitter.
\end{thebibliography}
\end{document}